# A HIGH-POWER 650 MHZ CW MAGNETRON TRANSMITTER FOR INTENSITY FRONTIER SUPERCONDUCTING ACCELERATORS

Grigory Kazakevich[#], Gene Flanagan, Rolland Johnson, Frank Marhauser, Michael Neubauer,
Muons, Inc., Batavia, 60510 IL, USA
Todd Treado, CPI, Beverly, 01915 MA, USA
Vyacheslav P. Yakovlev, Brian Chase, Sergei Nagaitsev, Ralph J. Pasquinelli,
Fermilab, Batavia, 60510 IL, USA

*Abstract*

A concept of a 650 MHz high-power magnetron transmitter with a fast control in phase and power, based on two-stage injection-locked CW magnetrons, has been proposed to drive Superconducting Cavities (SC) for intensity-frontier accelerators. The transmitter consists of two 2-stage magnetrons with outputs combined with a 3-dB hybrid to fulfil a fast control of output power required for SC accelerators. The output power is controlled by varying phase in the input of one of the 2-stage magnetrons. For output power up to 250 kW we expect the output/locking power ratio to be about 30 to 40 dB in CW or quasi-CW mode with long pulse duration. The phase control bandwidth in MHz range has been evaluated measuring transient time of the phase jump with a 2-stage CW magnetron model operating in pulsed regime. Description of the magnetron transmitter concept and it's modelling using frequency-locked commercial, CW, 1 kW, S-Band magnetrons operating in pulsed regime with long pulse duration is presented and discussed in this paper.

## INTRODUCTION

The state of the art superconducting linacs for proton and ion beams capable of accelerating protons and ions to several GeV with an average beam current of 1 to 10 mA or more are prospective for intensity frontier accelerator physics and are the basis for Accelerator Driven Systems (ADS), which have evoked significant interest the world over because of their capability to provide operation of sub-critical reactors in nuclear power stations, to incinerate the minor actinides and long-lived fission products of radiotoxic waste, and for potential utilization of Thorium as a nuclear fuel, [1, 2].

The high-intensity SC accelerators intended for non-relativistic or weakly relativistic particles require separate and independent feeding of each superconducting cavity with its own RF source controlled in phase and power in order to avoid growth of transverse and longitudinal emittances that may even result in beam loss. The RF source power, $P_a$, required for each SC accelerating a weakly relativistic proton beam with energy gain of $V_a$ can be estimated by:

$$P_a = V_a \cdot I_a.$$

Here $I_a$ is the accelerated current. Assuming that $V_a$=20 MeV, $I_a$ =10 mA one gets $P_a \approx$ 200 kW per the cavity.

Consequently, the costs of traditional high-power CW RF sources, such as TV klystrons or Inductive Output Tubes (IOTs), for separate feeding of the superconducting cavities of the GeV scale accelerators become prohibitive. The state of the art solid-state RF sources for 650 MHz at the required power are yet rather expensive as well. Hence the proposed concept of the magnetron transmitter with driving power of about 1 kW or less, being based on prototypes of commercial tubes looks as a solution potentially lowering costs of projects of the intensity frontier high-energy accelerators.

One of the primary concerns of using the magnetron transmitter to power superconducting cavities is the management of phase with a Low Level RF (LLRF) system generating a controlling signal which slowly varies phase on input of a 2-stage frequency-locked magnetron. The task is formally similar to one considering operation of a magnetron locked by slowly varying frequency/phase, [3-5]. This task was successfully solved theoretically considering the frequency-locked magnetron as a forced oscillator and proven experimentally [ibid]. Unlike Adler's approach [6], the developed technique analysing transient process in the locked magnetron allows numerical simulations of the frequency/phase deviations in time domain. Analysis of the simulations, successful experiments with the frequency-locked magnetrons, [3-8], and particularly our experimental results obtained with a frequency-locked 2-stage (cascade) magnetron model, [9] motivated our concept of the high power 650 MHz frequency-locked magnetron transmitter with a phase and power control.

First experimental modelling of the magnetron transmitter has been performed using CW 2.45 GHz magnetrons with power up to 1 kW, operating in a pulsed mode. The paper describes the experimental setup and the obtained results.

## EXPERIMENTAL FACILITY

The experimental facility was developed and built to study the operation of the CW single and two-stage frequency-locked magnetrons in pulsed regime with long pulse duration. The setup for the first experiments includes a pulsed modulator to feed the magnetrons, two single-stage magnetron modules with RF components, wide-band CW Travelling-Wave Tube (TWT) amplifier driven by a synthesizer, and measuring equipment, [9].

The modulator, [9], provides concurrent operation of two magnetrons with a controlled in the range of -39 deg

---

to 30 deg shift of the high voltage pulse relatively to the zero-crossing of the magnetron filament current allowing one to check the influence of the filament magnetic field on the magnetron phase stability.

## CONCEPT OF A MAGNETRON TRANSMITTER FOR SC ACCELERATORS

A concept of the magnetron transmitter (with phase and power to be controlled by a LLRF) is presented in Figure 1, [10].

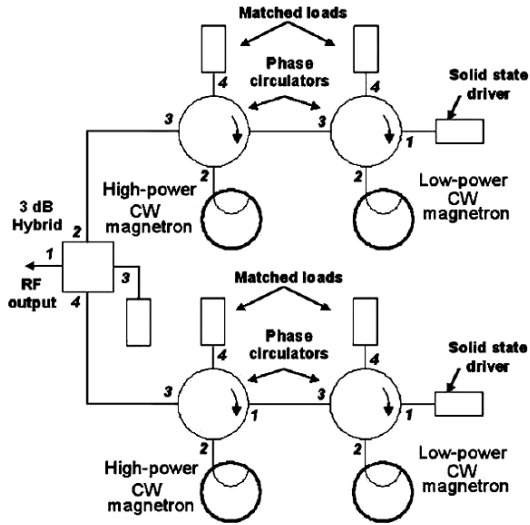

Figure 1: Conceptual scheme of the two-stage CW magnetron transmitter with a control in power and phase.

The conceptual model consists of two 2-stage magnetrons combined with, for example, a 3-dB $90^0$ hybrid. Vector sum of the output signals from each of the 2-stage magnetrons is intended to feed a SC. Rapid power modulation is achieved by varying the phase of the input signal of one of the 2-stage magnetrons relatively to the second one. For the proposed scheme one can estimate impact of difference of power of 2-stage magnetrons analysing amplitudes $U_1$ and $U_3$ on the ports #1 and #3 respectively. If the amplitudes ratio on ports #4 and #2 is $U_4/U_2 = \alpha$, one can write:

$$\frac{1}{\sqrt{2}}\begin{pmatrix} 0 & 1 & 0 & i \\ 1 & 0 & i & 0 \\ 0 & i & 0 & 1 \\ i & 0 & 1 & 0 \end{pmatrix} \cdot \begin{pmatrix} 0 \\ e^{i\varphi} \\ 0 \\ -i\alpha \cdot e^{i\varphi} \end{pmatrix} = \frac{1}{\sqrt{2}} \begin{pmatrix} (1+\alpha)e^{i\varphi} \\ 0 \\ i(1-\alpha)e^{i\varphi} \\ 0 \end{pmatrix} = \frac{1}{\sqrt{2}} \begin{pmatrix} U_1 e^{i\varphi} \\ 0 \\ U_3 e^{i\varphi} \\ 0 \end{pmatrix}.$$

Here the first matrix is S-matrix of the ideal hybrid, [11], the second one gives input voltages on the hybrid ports #2 and #4, the product gives waves on the ports #1 and # 3, respectively. Computation of power on ports #1 and #3 vs. ratio of power, $P_4/P_2$ on ports #4 and #2, respectively, plotted in Figure 2, was done using the obtained amplitude values. Figure 2 shows that noticeable difference in the 2-stage magnetrons power levels does not lead to significant loss of power in the high power magnetron transmitter which will provide a good efficiency, however the transmitter maximum output power is limited by a magnetron with lower power.

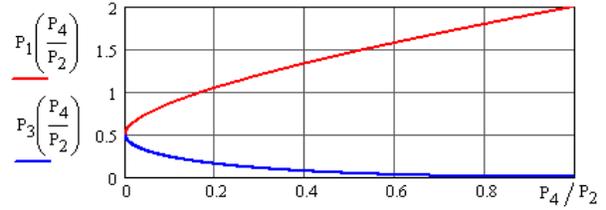

Figure 2: Computed power on the 3-dB $90^0$ hybrid output ports vs. ratio of power $P_4/P_2$ on it input ports.

## STUDY OF TWO-STAGE AND SINGLE-STAGE MAGNETRON OPERATION

A circuit diagram for the 2-stage (cascade) magnetron experiment is shown in Figure 3. The magnetrons were chosen to operate with the same locking frequency.

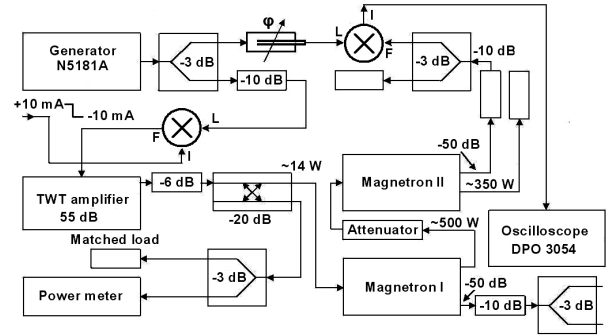

Figure 3: Experimental setup to study 2-stage magnetron.

The first magnetron was frequency-locked by the TWT amplifier; the second one was frequency-locked by the attenuated output signal of the first magnetron. Measured spectra in outputs of the both frequency-locked magnetrons at 5 ms pulse width are shown in Figure 4.

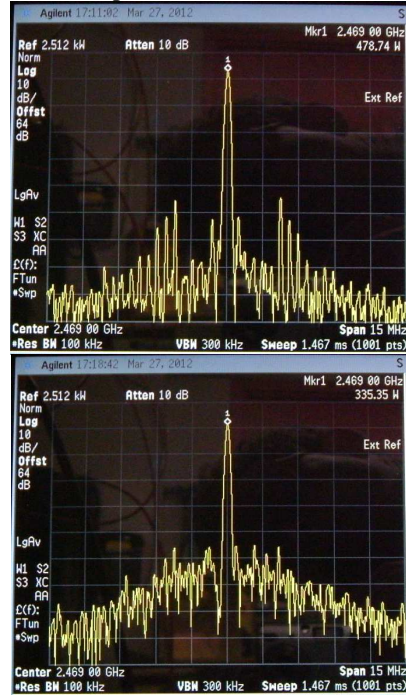

Figure 4: Spectra of the first (upper figure) and second (lower figure) magnetron output signals. The magnetrons are connected via a 20 dB attenuator.

Measured phase deviations of the 2-stage frequency-locked magnetron vs. the attenuator value are plotted in Figure 5. More detailed traces are presented in [9].

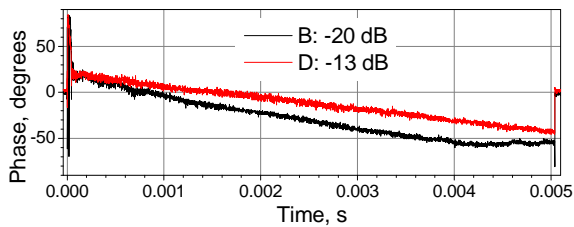

Figure 5: Phase deviations of the 2-stage frequency-locked magnetron vs. the attenuator between the stages.

Due to the perturbation of the forced oscillation coming from the pulse feeding of the first magnetron there is a finite time-to-lock ≤ 50 μs, [9]. The measured time-to-lock is adequate for intensity frontier accelerators with SC. To measure the response time of the 2-stage magnetron to a fast 180 degrees phase jump, we have used an additional mixer with an inverse connection, Figure 3. The response time of the 2-stage frequency-locked magnetron to the fast 180 degrees phase jump is plotted in Figure 6, showing that the bandwidth of a phase control for the 2-stage injection-locked magnetron is a few MHz.

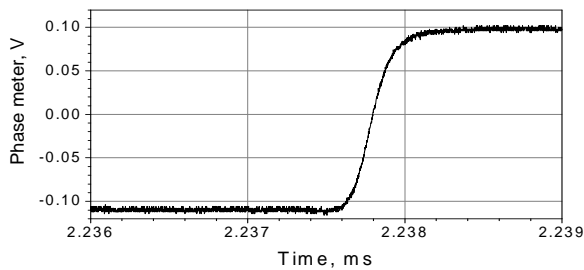

Figure 6: The frequency-locked 2-stage magnetron response on a fast 180 degrees phase jump.

Study of the influence of the magnetron filament current on the phase variation of the frequency-locked magnetron was performed with a single-stage magnetron with a pulse duration of 4.8 ms. We varied the modulator launch phase relatively to the zero-crossing of the magnetron filament current, which was ≈ 9 A. Figure 7 clearly shows that the filament magnetic field noticeably affects the phase stability of the magnetron. This affect has to be taken into account especially in pulsed operation of high-power magnetrons, in which the filament current is much higher.

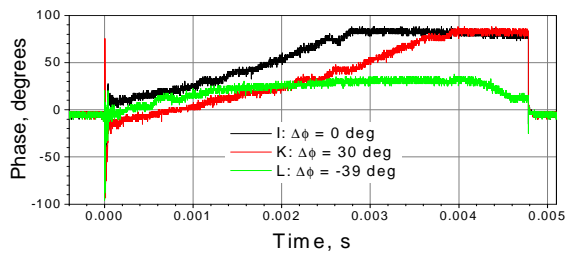

Figure 7: Single frequency-locked magnetron phase deviations vs. the filament phase.

The slow phase drift (trace L, where the phase detector is not saturated) is caused by two competing processes: an increase of the magnetron current due to pulsed overheating of the cathode caused by bombardment by returning electrons and a decrease of the current caused by an exponential drop of the magnetron voltage resulted from discharge of the storage capacitor. Both processes cause slow variations of the phase because of the frequency pushing effect in the magnetron.

Experimental check of the power variation with a 3-dB hybrid as it is shown in Figure 1 is not performed yet but the work is in progress. A model based on the 2.45 GHz CW commercial magnetrons is developed and currently built. Note that this technique is already well known in RF systems and there is no concern about its implementation, nevertheless we will check operation of the frequency locked magnetrons with a 3-dB combiner.

## CONCLUSION

A high-power magnetron transmitter consisting of two 2-stage frequency-locked magnetrons with a 3-db hybrid combiner has been proposed to drive superconducting cavities of the intensity frontier accelerators. A proof of principle has been performed modelling the 2-stage magnetron with the frequency-locked 2.45 GHz, CW, commercial magnetrons operating in pulse regime with pulse duration of 5 ms. Instantaneous phase noise less than few degrees was measured in operation of the 2-stage magnetron model with a peak-to-noise ratio in the magnetron spectrum at least of 45 dB. Time-to-lock measured for the 2-stage frequency-locked magnetron ≤ 50 μs was obtained at the ratio of output power to locking power of 33.5 dB. Measured time response of ≤ 200 ns on a fast phase jump proves ability of the 2-stage frequency-locked magnetrons to provide control in phase and power with a bandwidth up to few MHz. Results of the magnetron transmitter modelling are adequate for driving superconducting cavities. Further experiments in modelling of the proposed transmitter are in progress.